\documentclass[letterpaper,12pt]{article}
\usepackage{tabularx} 
\usepackage{amsmath}  
\usepackage{amssymb}
\usepackage{graphicx} 
\usepackage[margin=1in,letterpaper]{geometry} 
\usepackage{cite} 

\usepackage[final]{hyperref} 
\hypersetup{
	colorlinks=true,       
	linkcolor=blue,        
	citecolor=blue,        
	filecolor=magenta,     
	urlcolor=blue         
}
\usepackage{authblk}

\begin{document}

\title{Fibonacci topological phase in arrays of  anyonic chains}
\author{Hiromi Ebisu}
\affil{Department of Condensed Matter Physics,
Weizmann Institute of Science, Rehovot, Israel 76100}
\date{\today}
\maketitle

 \begin{abstract}


Fibonacci anyon, an exotic quasi-particle excitation, plays a pivotal role in realization of a quantum computer. Starting from a $SU(2)_4$ topological phase, in this paper we demonstrate a way to construct a Fibonacci topological phase which has only one non-trivial excitation described by the Fibonacci anyon. 
We show that arrays of anyonic chains created by excitations of the $SU(2)_4$ phase leads to the Fibonacci phase. We further demonstrate that our theoretical propositions can be extended to other topological phases.

\end{abstract}
\maketitle
  \section{Introduction}
Topologically order phases of matter, such as fractional quantum Hall states \cite{Tsui,laughlin1983anomalous} and spin liquid phases\cite{anderson1973resonating,Kalmeyer1987,wen1989chiral,read1991large,Kitaev2003}, have been studied intensively in the last decades. A beautiful aspect of these phases is that the phases are described by topological quantum field theory as the low energy effective theory~\cite{witten1989quantum}. Furthermore, the fascinating property of these phases is that they admit an exotic fractionalized excitation, namely anyon\cite{leinaas1977theory,Wilczek1982}. Non-Abelian anyon\cite{Frohlich1990,Moore1991,wen1991non}, whose braiding representation is described by a matrix in a degenerate ground state, is of particular importance from quantum information perspective. For practical purposes, the Fibonacci anyon\cite{Freedman2002,Freedman2003}, $\tau$ which is subject to the fusion rule $\tau\times\tau=I+\tau$ with vacuum being denoted by $I$, attracts a lot of interests as 
braiding these anyons yields a complete set of quantum operations in ground state manifold, 
which could be useful for quantum computations. Therefore, it would be desirable to obtain a topological phase which admits the Fibonacci anyon as an excitation.    \par 
There are several studies which construct the Fibonacci phase, \emph{a.k.a} the $(G_2)_1$ topological phase which habors only two types of excitations, vacuum $I$ and the Fibonacci anyon $\tau$. Mong \textit{et al} have shown that couplings of trenches of $\nu=2/3$ fractional quantum Hall edge modes generates the Fibonacci phase~\cite{Mong}. Hu and Kane\cite{Hu2018} have demonstrated that the Fibonacci phase arises in interacting $p$-wave superconductors. The goal of this paper is to present an alternative construction of the Fibonacci phase by introducing the $SU(2)_4$ topological phase and an arrays of nucleated anyonic excitations in the bulk. It turns out that couplings between these arrays drives the system to the desired Fibonacci phase. \par
Another motivation of this work is related to nucleation of a topological liquid. When we populate a topologically order phase with a set of non-Abelian anyons, it yields degeneracy in a ground state manifold. Interactions between the anyons lift the degeneracy, giving rise to a new collective ground state. 
It is known that one-dimensional (1D)\footnote{Throughout, we use the word dimension to mean the number of space dimension, not space-time dimension.} alignment of interacting non-Abelian anyons, that we call anyonic chain in this paper, realizes a non-trivial gapless collective mode~\cite{Feiguin2007,ludwig2011two}. 
It is interesting to investigate whether the interplay of interacting non-Abelian anyons forming in two-dimensional (2D) configuration allows us to obtain new kind of phases. 
Such a problem is generically hard to study with a few exceptions. 
For instance, it is possible to obtain the phase diagram for 2D vortex lattice of interacting Ising anyons (\emph{a.k.a.} Majorana zero modes) as demonstrated in Ref.~\cite{affleck2017majorana}.

However, it is still challenging to analyze phases in 2D configuration of interacting non-Abelian anyons beyond the Ising anyons. One useful way to tackle this issue is to take \textit{anisotropic} limit, meaning interaction between the anyons in horizontal direction is larger than the ones in the vertical direction. With this limit, we can resort to the formalism developed in Refs.~\cite{Kane2002,Teo2014}, where 2D topological phases are constructed by 1D conformal field theories (CFTs) arranged in parallel. 
As a simple example, consider $N$ copies of critical Majorana chain, $i.e.,$ $N$ copies of 1D free Majorana theories stack in parallel and introduce couplings between holomorphic (that is, right moving) field at a chain and anti-holomorphic (that is, left moving) field at the adjacent chain. The couplings gap out all the Majorana fields in the bulk but leave gapless chiral fields on the edges, indicating the system is equivalent to 2D topological superconductor ($p+ip$ superconductor)\footnote{Conversely, we can start with $N_x\times N_y$ lattice model of 2D topological superconductor and take the anisotropic limit. As the couplings in horizontal direction is greater than the ones in vertical direction, we can turn-off the vertical couplings for a moment, and focus on $N_y$ decoupled chains. Linearizing the spectrum of each chain, we obtain $N_y$ free Majorana theories. Now turn-on the vertical couplings. One can show that the model is topologically equivalent to $N_y$ Majorana theories with couplings between holomorphic and anti-holomorphic fields of adjacent chains, corroborating the legitimacy of the construction of the 2D topological superconductor from the 1D Majorana chains.  }. This logic can be applied to various topological phases such as Abelian and non-Abelian fractional quantum Hall states by considering networks of 1D CFTs where pairs of holomorphic and anti-holomorphic fields of adjacent chains are coupled. Furthermore, it was shown that a kink excitation in gapped area between adjacent chains coincides with an anyonic excitation in the bulk of the resulting 2D topological phase~\cite{Kane2002,Teo2014}.
Since this formalism allows us to map the 2D problem to 1D one, we can demonstrate that an arrays of anyonic chains would yield a new kind of topological phase, based on perturbed CFTs.


\par
The organization of this paper is as follows. In Sec.~\ref{model}, we present a way to construct the Fibonacci phase in the $SU(2)_4$ topological phase, consisting of three steps. In Sec.~\ref{ii}, we briefly comment on how
the logic developed in the preceding section to other cases of $SU(2)_k$. Finally, Sec.~\ref{conc} is devoted to conclusion and outlook. Technical details are relegated to Appendices. 
\section{Formulation}\label{model}
In this section, we demonstrate how the Fibonacci phase arises in the $SU(2)_4$ topological phase. There are three steps to realize it. 


\begin{figure}[h] 
        \centering \includegraphics[width=0.35\columnwidth]{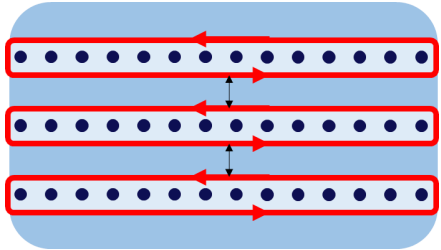}
        \caption{
                A schematic picture of the $SU(2)_4$ topological phase (blue rectangle). An anyonic chain consisting of $Z$ anyons depicted by dotted circles. The ``anti-ferromagnetic" state of this anyonic chain  yields the $\mathcal{M}(6,5)$ minimal conformal field theory (CFT) with central charge $c=4/5$ depicted by a red line. 
The coupling between adjacent anyonic chains is mediated by $X$ anyon (black arrow).        } \label{silver} 
\end{figure}

\subsection{\texorpdfstring{The $SU(2)_4$}{Lg} topological phase and anyonic chain}
 Before going to the details, let us first review the anyonic content of the $SU(2)_4$ topological phase. This phase is one of the chiral spin liquid phases which carries neutral chiral edge mode characterized by $SU(2)_4$ Wess-Zumino-Witten (WZW) conformal field theory (CFT) with central charge $c=2$. 
There are five types of anyons in this phase, labeled by $I$, $X$, $Y$, $Z$, $W$ in this paper. These five anyons are  
characterized by topological spin $\theta_{i}=e^{2\pi ih_i}$ ($i=I,X,Y,Z,W$) with 
$h_I=0$, $h_X=1/8$, $h_Y=1/3$, $h_Z=5/8$, $h_W=1$ (See also Appendix.~\ref{appendx}), which corresponds to five types of primary fields in the $SU(2)_4$ WZW CFT.

Consider a 1D alignment of anyonic excitations  
described by $Z$ anyons in the bulk such that fusion of each adjacent $Z$ anyon may occur, giving
\begin{equation}
    Z\times Z=I+Y.\label{p}
\end{equation}

\begin{figure}[h] 
        \centering \includegraphics[width=1\columnwidth]{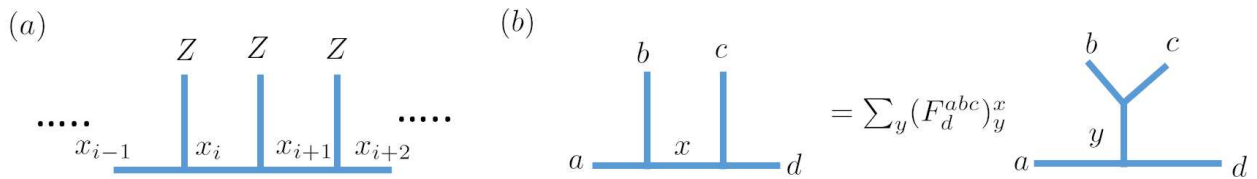}
        \caption{(a)~ A fusion diagram of the anyonic chain.
               (b)~Definition of the ``F-move''. The indices, $a,b,c,d,x,y$ are fusion channels.    } \label{fs} 
\end{figure}
To describe Hamiltonian of the anyonic chain (see Refs.~\cite{Feiguin2007,trebst2008short,buican2017anyonic} for more explanation of model construction), we introduce $N$ copies of $Z$ anyons, and a periodic fusion diagram given in Fig.~\ref{fs}(a), where $\{x_i\}\;(i=0,\cdots,N)$ with $x_0=x_N$ is allowed collection of fusion channels, $\emph{i.e.,}$ $x_{i+1}$ is determined from fusion between the $Z$ anyon and $x_i$. The Hilbert space of  the anyonic chain is defined by $\left|x_0,\cdots,x_{N-1}\right>$ which is in one-to-one correspondence with admissible fusion diagram given in Fig.~\ref{fs}(a). We write Hamiltonian of the $Z$ anyonic chains as
\begin{equation}
    H=-J\sum_iP_i^{Z,I},\label{pro}
\end{equation}
where $P_i^{Z,I}$ is the projection operator which sends the outcome of the fusion of adjacent $Z$ anyons to $I$. Using the F-move defined in Fig.~2(b), the action of $P_i^{Z,I}$ on a state is given by 
\begin{eqnarray}
    P_i^{Z,I}\left|x_0,\cdots,x_{i-1},x_{i},x_{i+1},\cdots,x_{N-1}\right>\nonumber\\
    =\sum_{x^{\prime}_i}(F^{x_{i-1}ZZ}_{x_{i+1}})_Z^{x_i}(F^{-1 x_{i-1}ZZ}_{x_{i+1}})_{x_i^{\prime}}^{Z}\left|x_0,\cdots,x_{i-1},x_{i},x_{i+1},\cdots,x_{N-1}\right>.
\end{eqnarray}
The coefficient in the matrix form of the F-move is determined from the analysis of quantum group. We don`t show it here, as it is not necessary in subsequent discussion. When $J>0$, the fusion outcome of the adjacent $Z$ anyons is energetically favored to be $I$. 
\par
The chain of this $Z$ anyons is reminiscent of a $1/2$-spin chain with nearest neighbors coupling. 
Similarly to the fact that there are two orders in the spin chain, namely, anti-ferromagnetic/ferromagnetic state which favors spin singlet/triplet state in each bond depending on the sign of the coupling, the anyonic chain admits two different orders as studied by
Refs.~\cite{Feiguin2007,ludwig2011two}. 
It is known that if the chain of the $Z$-anyons is ``anti-ferromagnetic", meaning fusion outcome $I$ is energetically favored, the anyon chain yields a silver of a gapped nucleated liquid carrying 
a gapless edge mode at the interface with the $SU(2)_4$ topological phase.  The gapless theory is decried by the minimal CFT, $\mathcal{M}(6,5)$ with
central charge $c=4/5$, the same universality class as the tetracritical Ising model.
\par
We envisage array of the critical $Z$-anyon chains (with the index $j$ to denote the $j$th chain) in the bulk of the $SU(2)_4$ topological phase, each of which is described by $1+1$-dimensional $c=4/5$ CFT. We would like to couple these arrays by introducing a weak tunneling term between adjacent chains. 
To this end, 
we assume that the silver of a gapped liquid nucleated by the anyonic chain has finite weight~\cite{ludwig2011two} so that the right and left moving gapless modes are spatially separated as shown in Fig.~\ref{silver}. Therefore, we focus on the coupling between the closer counter-propagating edge modes of the adjacent chains, as indicated by black arrows in Fig.~\ref{silver}. More concretely, in this section, we assume that the closer counter-propagating edge modes are described by the holomorphic and anti-holomorphic sectors at $j$th and $j+1$th chains, respectively. 
Further, we
assume that the distance between the adjacent
anyonic chains is short so that a  tunneling between the chains may occur.
In this process, it is important that what kind of topological fluid is filled between the anyonic chains. In analogy with Ref.~\cite{Mong}, which discusses that a tunneling of $\mathbb{Z}_3$ parafermions between the critical chains of three states Potts model is allowed as $\mathbb{Z}_3$ topological phase is filled between the chains, we claim that whether
a form of a tunneling between the adjacent chains consisting of primary fields of CFT of the chain is allowed or not depends on whether there is a anyonic excitation in the topological phase between the chains has the same conformal weight as the primary field.

Focusing on a pair of adjacent chains, \emph{e.g.,} the $j$th and $j+1$th anyonic chains, 
we consider following form of tunneling between the chains:
\begin{equation}
   \phi_{1,2}^{j}\overline{\phi}_{1,2}^{j+1},\label{pp}
\end{equation}
where $\phi_{r,s}^j$ represents a primary field with index $(r,s)$ in holomorphic sector in the $\mathcal{M}(6,5)$ minimal model at $j$th chain 
with conformal weight (see also Table.~\ref{tb})
\begin{equation}
  h_{r,s} = \frac{(6r-5s)^2-1}{120}=h_{5-r,6-s},\;\;1\leq r\leq 4,1\leq s\leq 5.\label{mini}
\end{equation}
\begin{table}
\begin{center}
\begin{tabular}{ c|c c ccc}
$4$&$3$ & $13/8$ & $2/3$&$1/8$&$0$ \\ $3$&$7/5$ & $21/40$ & $1/15$&$1/40$& $2/5$ \\ 
 $2$& $2/5$ & $1/40$ & $1/15$&$21/40$&$7/5$ \\ $1$&$0$ & $1/8$ & $2/3$&$13/8$& $3$ \\
  \hline\\
  $r/s$&$1$&$2$&$3$&$4$&$5$
\end{tabular}
\end{center}
\caption{Conformal weight in the $\mathcal{M}(6,5)$ minimal model according to Eq.~(\ref{mini}). Notice that due to the relation $h_{r,s}=h_{5-r,6-s}$, there are ten distinct primary fields. }\label{tb}
\end{table}
Likewise, $\overline{\phi}_{r,s}^{j+1}$ denotes the primary field in anti-holomorphic sector at $j+1$th chain with the conformal weight $h_{r,s}$.
According to the claim  above Eq.~(\ref{pp}), the tunnelling, $\phi_{1,2}^{j}\overline{\phi}_{1,2}^{j+1}$, is allowed as in the bulk there is an anyonic excitation whose conformal weight is the same as the primary field $\phi_{1,2}^{j}$ and $\phi_{1,2}^{j+1}$.

\subsection{\texorpdfstring{Perturbed minimal model: $\Phi_{(1,2)}$}{Lg} deformation}

The holomorphic sector at $j$th chain and the anti-holomorphic sector at $j+1$th chain constitute the $1+1$-dimensional minimal CFT. 
With the tunneling that we have considered in the preceding subsection, the theory is described by
\begin{equation}
S=S^{\text{CFT}}_{\mathcal{M}(6,5)}+\lambda\int dzd\bar{z}\Phi_{(1,2)}(z,\bar{z}
)\label{theory},
\end{equation}
where $z$ and $\overline{z}$ are complex coordinate which originates from the one in $1+1$-dimension, $i.e.,$ $(t,x)$, a non-chiral primary field is defined by $\Phi_{(1,2)}=\phi^j_{1,2}\overline{\phi}^{j+1}_{(1,2)}$ (other primary fields are similarly defined), and $\lambda$ is dimensional constant which scales as $\lambda\sim aM^{[2-2h_{(1,2)}]}=aM^{3/2}$ with $a$ and $M$ being dimensionless constant and mass scale, respectively. As mentioned in the introduction, we take the anisotropic limit, hence, we require $J\gg |\lambda|$, where $J$ appeared in Eq.~(\ref{pro}).

\par
The theory~(\ref{theory}) is known as the $\Phi_{(1,2)}$ deformation of the minimal model, which is studied in view of integrability and quantum group~\cite{smirnov1991exact,mussardo1992off}.
Based on these analysis, it turns out that the theory~(\ref{theory}) describes a gapped fluid with two fold ground state degeneracy  admitting one type of kink excitation, which is associated with the Fibonacci anyon. 
Indeed, due to the fusion rule
\begin{equation}
    [\Phi_{(1,2)}]\times[\Phi_{(r,s)}]=[\Phi_{(r,s-1)}]+[\Phi_{(r,s+1)}],
\end{equation}
the $\Phi_{(1,2)}$ perturbation mixes the states generated by $\Phi_{(r,s)}$ and those by $\Phi_{(r,s\pm1)}$. It follows that the Hilbert space in this theory splits into two sectors: $\mathcal{H}=\mathcal{H}_1+\mathcal{H}_2$ with
\begin{equation}
    \mathcal{H}_r=\bigoplus_{s=1}^{5}\mathcal{V}_{(r,s)}\otimes \overline{\mathcal{V}}_{(r,s)},\;\;r=1,2.\label{hitagi}
\end{equation}
Here, $\mathcal{V}_{(r,s)}[\overline{\mathcal{V}}_{(r,s)}]$ is the irreducible representation of the Virasoro algebra in the holomorphic [anti-holomorphic] sector with highest weight $h_{r,s}$. Irrespective of the sign of $\lambda$ in Eq.~(\ref{theory}), the theory becomes gapped with two fold ground state degeneracy, associated to the field $\Phi_{(r,r)}$ in the sector $\mathcal{H}_r$ ($r=1,2$). Furthermore, we can associate these two ground states to spin-$0$ and $1$ representations of the $U_q[sl(2)]$ quantum group symmetry which are denoted by $\left|0\right>$ and $\left|1\right>$, and an elementary kink excitation which interpolates between the two vacua 
is also represented by spin-$1$ representation of the same quantum group symmetry. The fusion rule of this spin-$1$ representation is given by (Appendix.~\ref{quantum group})
\begin{equation}
1\times1=0+1,\;\; 0\times1=1,\label{fib22}
\end{equation}
which coincides with the one of the Fibonacci anyon.

\subsection{Hilbert space of the anyonic chain}
In the previous subsection, we have seen that each pair of holomorphic/anti-holomorphic sectors of the adjacent anionic chains constitutes a 1D gapped fluid which admits the Fibonacci kink. We will see that array of such a gapped 1D fluid gives rise to the desired Fibonacci topological phase. To do so, we resort to the technique developed in Ref.~\cite{Mong} and investigate how the form of the Hilbert space of each anyonic chain put a constraint on the number of ground states in the bulk.\par 
As we have discussed, the adjacent anyonic chains are coupled via tunneling of the excitation corresponding to a primary field with index $(r,s)=(1,2)$. We envisage that the bulk topological phase is placed on a torus geometry by imposing periodic boundary condition along the direction perpendicular to the 1D anyonic chains. In the present case, this is done by introducing the tunneling in the form of Eq.~(\ref{pp}) with the superscript $j$ and $j+1$ being replaced with $N$ and $1$ ($N$ is the number of the anyonic chains).
Focusing on one anyonic chain, the tunneling process is effectively regarded as transfer of the excitation with index $(r,s)=(1,2)$ \emph{within the same chain}; at the $j$th chain, the excitation with $(r,s)=(1,2)$ comes in from the holomorphic sector of the $j-1$th chain and the same type of the excitation gets out from the anti-holomorphic sector of the $j$th chain to the $j+1$th chain.   
Therefore, the Hilbert space of the $j$th chain is given by
\begin{equation}
    \mathcal{H}^j=\bigoplus_{\substack{(r,s) \\ (t,u)}}\mathcal{V}^j_{(r,s)}\otimes \overline{\mathcal{V}}^j_{(t,u)},\label{jj}
\end{equation}
where the index $(r,s)$ and $(t,u)$ are related by fusion with primary field with index $(1,2)$. 
For instance, $(r,s)=(1,1)$ and $(t,u)=(1,2)$ is allowed combination due to the fusion rule $[\Phi_{(1,2)}]\times[\Phi_{(1,1)}]=[\Phi_{(1,2)}]$. We derive the same form of the Hilbert space from another perspective calculating twisted partition function, see Appendix.~\ref{boundary}.
\par
Consider three anyonic chains, the $j-1$th, $j$th, and $j+1$th chains. From the discussion in the previous subsection, between the $j-1$th and $j$th chains as well as the $j$th and $j+1$th chains, we have gapped theories each of which has two fold ground state degeneracy. Naively, one expects that such gapped theories contribute to four fold degeneracy, which, however, turns out to be incorrect; two states are eliminated, giving two fold ground state degeneracy. To see why, we denote multiplication of the ground states in areas between the $j-1$th and $j$th chains and the ones between $j$th and $j+1$th chains by $\left|p\right>^{j-1/2}\left|q\right>^{j+1/2}$ ($p,q=0,1$). 
Suppose  $\left|0\right>^{j-1/2}\left|1\right>^{j+1/2}$ is allowed ground state.
From Eq.~(\ref{hitagi}), the state $\left|0\right>^{j-1/2}$ belongs to the Hilbert space
\begin{equation}
      \mathcal{H}_1^{j-1/2}=\bigoplus_{s=1}^{5}\mathcal{V}_{(1,s)}^{j}\otimes \overline{\mathcal{V}}_{(1,s)}^{j-1},\label{hanekawa}
\end{equation}
where we have introduced the superscript to denote the anyonic chain. Similarly, the state $\left|1\right>^{j+1/2}$ belongs to the Hilbert space
\begin{equation}
      \mathcal{H}_2^{j+1/2}=\bigoplus_{s=1}^{5}\mathcal{V}_{(2,s)}^{j+1}\otimes \overline{\mathcal{V}}_{(2,s)}^{j}.\label{tsubasa}
\end{equation}
Eqs.~(\ref{hanekawa})(\ref{tsubasa}) implies that the state in question, $\left|0\right>^{j-1/2}\left|1\right>^{j+1/2}$ forces the Hilbert space of the $j$th anyonic chain to have the form $\sim \bigoplus_{s,s^{\prime}}^5\mathcal{V}^j_{(1,s)}\otimes \overline{\mathcal{V}}^j_{(2,s^{\prime})}$. However, this from is not allowed due to Eq.~(\ref{jj}). Indeed, $\mathcal{V}^j_{(1,s)}$ and $\overline{\mathcal{V}}^j_{(2,s^{\prime})}$ is not related by fusion with the primary field of the index $(1,2)$. Therefore, the state 
$\left|0\right>^{j-1/2}\left|1\right>^{j+1/2}$ is not allowed ground state. The similar argument shows that 
$\left|1\right>^{j-1/2}\left|0\right>^{j+1/2}$ is also excluded, concluding that there are two ground states, $\left|p\right>^{j-1/2}\left|p\right>^{j+1/2}$ ($p=0,1$).\par
The arrays of anyonic chains with weak tunnelings between adjacent chains yields a 2D gapped phase 
in the same spirit as the coupled wire construction~\cite{Kane2002,Teo2014}. The iterative use of the argument on the form of the Hilbert space given above shows that the arrays of the anyonic chains produces a gapped phase
which has two ground state degeneracy in the torus geometry, admitting the Fibonacci anyon as an excitation. Due to the well known fact that the number of the ground state degeneracy of a topological phase in a torus geometry is equivalent to the number of types of anyons (\emph{i.e.,} superselection sectors, in the present case, vacuum $I$ and the Fibonacci anyon $\tau$),
and that a kink excitation in gapped area between adjacent chain, in the present case kink excitation subject to the fusion rules in Eq.~(\ref{fib22}), coincides with the anyonic excitation in the bulk of the 2D topological phase~\cite{Kane2002,Teo2014}, 
the phase we have obtained is the desired Fibonacci phase, as advertised.\par
Further evidence that the resulting phase is the Fibonacci topological phase might be obtained by investigating topological entanglement entropy in a manner akin to Ref.~\cite{Mong}, from which one can obtain a total quantum dimension. To implement this analysis, we have to rely on numerical calculations (truncated conformal space approach), which hopefully leave for future works.
\section{Brief comments on other cases}\label{ii} The logic presented in the previous section, consisting of three steps, can be easily applied to other cases of the $SU(2)_k$ topological order phase. 
We summarize these three steps as follows:
\begin{enumerate}
    \item 
    Make an alignment of non-Abelian anyons corresponding to the $\phi_{\frac{k-1}{2}}$ primary field with fusion rule $\phi_{\frac{k-1}{2}}\times \phi_{\frac{k-1}{2}}=\phi_0+\phi_1$ in the $SU(2)_k$ topological phase. Assuming that the anyonic chain is in the ``anti-ferromagnetic" state, \textit{i.e.,} the outcome of fusion of the adjacent anyons it  projected  to  the  sector  with  fusion  channel $\phi_0$, then the anyonic chain leads to a gapless collective mode. This gapless mode is characterized by~$\mathcal{M}(k+2,k+1)$ minimal CFT. Find a relevant tunneling operator between adjacent anyonic chains. The tunneling term consists of a pair of primary fields in the holomorphic anti-holomorphic sector in the adjacent chains. We require that the primary field carry the same conformal spin~(mod $1$) as the one of an anyon in the bulk phase.
    \item 
    If you can successfully find an appropriate tunneling term, then the CFT consisting of a pair of holomorphic/anti-holomorphic sectors in the adjacent chains is perturbed by this tunneling term. The theory leads to 1D gapped fluid with non-trivial ground state degeneracy harboring kink excitations subject to the fusion rule determined by the quantum group symmetry.  
    \item Due to the analysis of the form of the Hilbert space in each anyonic chains, one can show that arrays of the anyonic chains with the tunneling between the adjacent chains considered in the previous steps manifests the 2D topological phase with the same ground state degeneracy as the one of the 1D gapped fluid in step~2 in a torus geometry. Furthermore, the phase admits anyons with the same fusion rule as the one of the spin representation of kink excitations in the gapped 1D fluid.
\end{enumerate}
As an example, we can apply these three steps to the $SU(2)_3$ topological phase. The gapless collective mode of ``anti-ferromagnetic" anyonic chain is characterized by the $\mathcal{M}(5,4)$ minimal CFT with central charge $c=7/10$ (\emph{a.k.a.} tricritical Ising CFT). Through a bulk anyon excitation corresponding to the primary field $\phi_1$, adjacent anyonic chains are coupled, which corresponds to the $\Phi_{1,3}$ perturbation of the $c=7/10$ CFT;
a 1D theory consisting of a pair of holomorphic/anti-holomorphic sectors of adjacent anyonic chains is described by
\begin{equation}
S=S^{\text{CFT}}_{\mathcal{M}(5,4)}+\lambda\int dzd\bar{z}\Phi_{(1,3)}(z,\bar{z}
)\label{theory2}.
\end{equation}
When $\lambda<0$ it is known that the theory is gapped with three fold degeneracy. The fundamental kink interpolating between the three vacua is described by spin-$1/2$  representation with fusion rule
\begin{equation}
    \frac{1}{2}\times\frac{1}{2}=0+1,
\end{equation}
reminiscent of the fusion rule of the Ising anyon (fusion rule of the spin-$1$ representation on the right hand side is given by $1\times 1=0$). Therefore, we obtain the 2D Ising topological phase by coupling of the anyonic chains. \par Another example is the $SU(2)_5$ topological phase. The anti-ferromagnetic state of the anyonic chains is characterized by $\mathcal{M}(7,6)$ minimal CFT with central charge $c=6/7$. The adjacent anyonic chains are coupled via the excitation corresponding to the primary field $\phi_{5/2}$ in the bulk of the $SU(2)_5$ topological phase, giving the $\Phi_{1,2}$ perturbation in the $c=6/7$ CFT. Similarly to the previous example, a 1D theory consisting of a pair of holomorphic/anti-holomorphic sectors of adjacent anyonic chains is described by
\begin{equation}
S=S^{\text{CFT}}_{\mathcal{M}(7,6)}+\lambda\int dzd\bar{z}\Phi_{(1,2)}(z,\bar{z}
)\label{theory23}.
\end{equation}
It is known that the perturbation with $\lambda<0$ leads to a gapped theory with three fold ground state degeneracy. A fundamental kink excitation interpolating between the vacua is given by spin-$1$ representation, subject to the fusion rule
\begin{equation}
    1\times 1=0+1+2.
\end{equation}
with $1\times2=1, 2\times2=0$. These fusion rules coincide with the ones of the $SO(3)_5$ topological phase. Therefore, coupling of the anyonic chains leads to the $SO(3)_5$ topological phase. It is worth mentioning that braiding representations of the  excitations of this phase constitute a complete set of quantum operation~\cite{rowell2009classification}, similarly to the Fibonacci phase.\par
As an aside, the adjacent anti-ferromagnetic chains in the $SU(2)_4$ topological phase can be coupled via bulk excitation of the $Y$-anyon, giving the $\Phi_{1,3}$ deformation of the $\mathcal{M}(6,5)$ minimal model. When the sign of the  couplings is negative, one obtains the $SU(2)_3$ topological phase.  
\section{Conclusions and outlook}\label{conc}
In this work, we present a construction of the Fibonacci phase in array of anyonic chains in the $SU(2)_4$ topological phase. Combination of the analysis of the perturbed CFT and the form of the Hilbert space of each anyonic chain allows us to obtain the Fibonacci phase, which harbors the Fibonacci anyon as the only non-trivial excitation. The way we present in this work is applicable to other cases. Before closing this section, we give several comments on future directions. \par
For realization purposes, it is important to find  Hamiltonian formalism to describe our phases. However, this task is challenging up to now. Especially, much is not known about how to write explicitly a concrete model to realize the anti-ferromagnetic anyonic chains, \emph{i.e.,} a model Hamiltonian which projects to sector $I$ when two adjacent anyons are fused as demonstrated in Eq.~(\ref{p}). One useful direction to address this issue is to introduce a $SU(2)$ current-current interaction in a fermionic system. As illustrated by Ref.~\cite{tsvelik2014integrable}, one can 
rewrite the interaction by parafermionic mass term where the mass is accompanied with a bosonic field. By adding a term which couples with the bosonic field, the spatial modulation of the mass of the parafermionic field may bind zero modes which are regarded as non-Abelian anyons, leading to an interacting anyonic chain. It turns out that one obtains a
collective gapless state corresponding to the ferromagnetic anyonic chain~\cite{tsvelik2014integrable,borcherding2018signatures}. It would be an interesting direction to investigate whether our construction is described by  Hamiltonian in the similar fashion. \par It would also be interesting to find a possibility to get a gapless theory by different means from critical anyonic chains. For instance, we prepare many copies of 2D topological phases with finite size carrying a chiral edge mode characterized by a CFT corresponding to the bulk topological phase. We place them consecutively. Focusing on a pair of adjacent topological phases, these two phases are separated by a pair of counter-propagating edge modes. Interactions or couplings between these modes may lead to a different gapless mode from the original edge mode. Systematic way to investigate the existence of such a gapless mode would be resorting to a recent work~\cite{ji2019noninvertible} which provides a way to find a partition function of the possible gapless mode of a given topological phase, based on bootstrap analysis. These new gapless modes may be coupled via an excitation of the bulk of the topological phase, similarly to our construction. Since the gapless CFT is different from the original edge modes, one expects that coupling of the CFTs would yield a new phase which is different from the original 2D topological phase. 

\section*{Acknowledgment} The author thanks Biswarup Ash, Rohit R. Kalloor, Yuval Oreg for helpful discussion. 
This work is supported by the Koshland postdoc fellowship and partially by  the ERC under the European Union’s Horizon 2020 research and innovation programme (grant agreement LEGOTOP No 788715).
\appendix
\section{\texorpdfstring{$SU(2)_4$}{Lg} WZW CFT}\label{appendx}
There are $k+1$ primaries of the $SU(2)_k$ WZNW CFT labeled by $\phi_i$ ($i=0,1/2,\cdots,k/2$) whose conformal weights are given by $i(i+1)/(k+2)$. The fusion rules of these primaries are
given by~\cite{DiFrancesco1997}
\begin{equation}
\phi_i\cdot\phi_j=\sum_{\substack{l=|i-j| \\ l-|i-j|\in\mathbb{Z}}}^{\text{min}(i+j,k-(i+j))}\phi_l\label{fusion}.
\end{equation}
In the case of $k=4$, there are five primary fields, $\phi_0,\cdots,\phi_4$\ with conformal weight being $0$, $1/8$, $1/3$, $5/8$, $1$ which is labed by $I,X,Y,Z,W$, respectively. Referring to Eq~(\ref{fusion}), we summarize the fusion rules of these primaries in Table.~\ref{fs}. 
\begin{table}[h]
\begin{center}
\begin{tabular}{ c|c c cc}
&$X$&$Y$&$Z$&$W$\\\hline$X$&$I+Y$&&&\\$Y$&$X+Z$&$I+Y+W$&&\\
$Z$&$Y+W$&$X+Z$&$I+Y$&\\
$W$&$Z$&$Y$&$X$&$I$
\end{tabular}
\end{center}
\caption{Fusion rules of the primary fields in the $SU(2)_4$ WZW CFT. The  missing  half  of  the  table may be filled in by commutativity. }\label{fs}
\end{table}
\section{Quick review on quantum group}\label{quantum group}
The quantum group $SL_q(2)$~\cite{majid2000foundations}, which is the deformation of the algebra of functions over $SL(2)$. The universal enveloping algebra $U_q[sl(2)]$ is generated by $\{H,J_{+},J_{-}\}$ satisfying the relation 

\begin{equation}
    [J_{+},J_{-}]=\frac{q^H-q^{-H}}{q-q^{-1}},\;[H,J_{\pm}]=\pm2J_{\pm}.\label{mnk}
\end{equation}
The irreducible representations of $U_q[sl(2)]$ are labelled by $j=0,\frac{1}{2},1,\cdots,$ acting on a Hilbert space $V_j$ with basis vector $\left|j,m\right>$
$(-j\leq m\leq j)$
 as follows:
\begin{equation}
    J_3\left|j,m\right>=m\left|j,m\right>,\;J_{\pm}\left|j,m\right>=\sqrt{[j\mp m]_q[j\pm m+1 ]_q}\left|j,m\right>\label{us}
\end{equation} 
with $[n]_q=\frac{q^n-q^{-n}}{q-q^{-1}}$. When $q\to 1$, the commutation relation~(\ref{mnk}) and the eigenvalue in Eq.~(\ref{us}) reduces to the ones of the standard $SL(2)$ group, therefore the quantum group $SL_q(2)$ becomes the $SL(2)$ group. \par
 
When $q$ is not a root of unity, the irreducible representations have dimension $2j+1$. On the contrary, when $q$ is a root of unity, the Hilbert space is truncated so that the allowed spins are $\{0,1/2,\cdots,j_{\max}\}$
with $j_{max}=N/2-1$ such that $q^N=\pm1$. Accordingly, introducing $a,b,c$ which are spins of this representation, tensor product of these spins is given by
\begin{equation}
    a\times b=\sum_{c=|a-b|}^{\min\{a+b,\; 2j_{max}-a-b\}}c.
\end{equation}
When minimal CFT $\mathcal{M}(m+1,m)$ is perturbed by $\lambda\Phi_{(1,3)}$, $(\lambda<0)$, the theory is gapped and the kink interpolating between adjacent vacua is described  by spin-$1/2$ representation of the quantum group $SL_q(2)$ with $q=e^{i\pi\frac{m+1}{m}}$. In the case of the minimal CFT $\mathcal{M}(m+1,m)$ being perturbed by $\lambda\Phi_{(1,2)}$, the fundamental kink excitation which interpolates between vacua is described by spin-$1$ representation of the same quantum group. 

\section{Twisted partition function}\label{boundary}
We present an alternative way to derive the form of the Hilbert space at $j$th anyonic chain given in Eq.~(\ref{jj}).\par
In Sec.~\ref{model}, we consider  gapless anyonic chains characterized by minimal CFT, $\mathcal{M}(6,5)$ with
central charge $c=4/5$,  coupled by $\Phi_{1,2}$, namely, primary field with index $(1,2)$. Focusing on one anyonic chain, we would like to find the Hilbert space of the chain. To do so, we take into account the presence of the tunneling of the primary field $\Phi_{1,2}$ properly. Suppose we put the theory on a geometry of a cylinder, $S^1\times \mathbb{R}$, where spatial direction is along the circle. 
The tunneling term at a local point of $x$ is visualized by a line along the time $\mathbb{R}$ direction, tempting us to interpret the tunneling term as the topological defect line which modifies the quantization by a twisted periodic boundary condition.  
We claim that the Hilbert space in question can be read from 
a twisted partition function with topological defect line corresponding to the primary field $\Phi_{1,2}$ being inserted along the time direction.\par
Let us briefly recall the defining  property of the topological defect line~\cite{frohlich2007duality,davydov2011invertible}. All the physical observable are invariant under continuous deformation of the topological defect lines. They are denoted by $\mathcal{L}_g$ where $g$ is global symmetry operation.
When a state $\left|\phi_k\right>$ crosses the $\mathcal{L}_g$, the symmetry $g$ is acted on the state. More rigorously, the topological defect line is described by the Verlinde line~\cite{verlinde1988fusion,petkova2001generalised} which acts on a state $\left|\phi_k\right>$ by
\begin{equation}
\mathcal{L}_g\left|\phi_k\right>=\frac{S_{gk}}{S_{0k}}\left|\phi_k\right>
\end{equation}
with $S_{gk}$ is modular $S$-matrix of the CFT.\par
In the present case, 
for the $(1,2)$ primary field, we can defined 
the Verlinde line $\mathcal{L}_{(1,2)}$ via
\begin{equation}
\mathcal{L}_{(1,2)}\left|\phi_{(r,s)}\right>=\frac{S_{(1,2)(r,s)}}{S_{(1,1)(r,s)}}\left|\phi_{(r,s)}\right>,
\end{equation}
where $\left|\phi_{(r,s)}\right>$ is a state corresponding to the ($r,s$) primary field in the  $c=4/5$ CFT.
The twisted partition function is calculated as follows. We start with the diagonal partition function\footnote{Depending on the microscopic details of the individual chain, there could be a possibility that the partition function of the  individual chain without tunnelling has the form of $\mathbb{Z}_2$ twisted one, $\sum_{(r,s),(t,u)}N_{(1,5),(r,s)}^{(u,t)}\overline{\chi}_{(r,s)}\chi_{(t,u)}$, instead of diagonal one~(\ref{dg}). Analogously to the case of the ``golden chain"~\cite{Feiguin2007}, we speculate that the partition function of each chain has diagonal one ($\mathbb{Z}_2$ twisted one)   when the chain length is even (odd), therefore we set the chain length to be even so that partition function of each chain has the diagonal form~(\ref{dg}).    }
\begin{equation}
    Z_{I,I}=\sum_{(r,s)}\overline{\chi}_{(r,s)\chi_{(r,s)}}\label{dg}
\end{equation}
and consider inserting the topological defect line along the spacial direction. Then the partition function is the form of
\begin{equation}
    Z_{I,(1,2)}=\sum_{(r,s)}\frac{S_{(1,2)(r,s)}}{S_{(1,1)(r,s)}}\overline{\chi}_{(r,s)}\chi_{(r,s)},
\end{equation}
where $\chi_{(r,s)}[\overline{\chi}_{(r,s)}]$ is the character corresponding to the $(r,s)$ primary field in the holomorphic [anti-holomorphic] sector of the $c=4/5$ CFT.
Implementing $S$-transformation
and using the  Verlinde formula~\cite{verlinde1988fusion},
one finds the desired twisted partition function:
\begin{equation}
    Z_{(1,2),I}=\sum_{(r,s),(t,u)}N_{(1,2),(r,s)}^{(u,t)}\overline{\chi}_{(r,s)}\chi_{(t,u)}\label{twis}
\end{equation}
with $N_{(1,2),(r,s)}^{(u,t)}$ being the fusion coefficient of the $c=4/5$ CFT. 
From the form of the twisted partition function~(\ref{twis}), we obtain the Hilbert space given in Eq.~(\ref{jj}).





\bibliographystyle{ieeetr}
\bibliography{ref}

\begin{thebibliography}{10}

\bibitem{Tsui}
D.~C. Tsui, H.~L. Stormer, and A.~C. Gossard {\em Phys. Rev. Lett.}, vol.~48,
  pp.~1559--1562, 1982.

\bibitem{laughlin1983anomalous}
R.~B. Laughlin, ``Anomalous quantum hall effect: an incompressible quantum
  fluid with fractionally charged excitations,'' {\em Physical Review Letters},
  vol.~50, no.~18, p.~1395, 1983.

\bibitem{anderson1973resonating}
P.~W. Anderson, ``Resonating valence bonds: A new kind of insulator?,'' {\em
  Materials Research Bulletin}, vol.~8, no.~2, pp.~153--160, 1973.

\bibitem{Kalmeyer1987}
V.~Kalmeyer and R.~B. Laughlin {\em Phys. Rev. Lett.}, vol.~59, pp.~2095--2098,
  1987.

\bibitem{wen1989chiral}
X.-G. Wen, F.~Wilczek, and A.~Zee, ``Chiral spin states and
  superconductivity,'' {\em Physical Review B}, vol.~39, no.~16, p.~11413,
  1989.

\bibitem{read1991large}
N.~Read and S.~Sachdev, ``Large-n expansion for frustrated quantum
  antiferromagnets,'' {\em Physical review letters}, vol.~66, no.~13, p.~1773,
  1991.

\bibitem{Kitaev2003}
A.~Kitaev {\em Annals of Physics}, vol.~303, no.~1, pp.~2 -- 30, 2003.

\bibitem{witten1989quantum}
E.~Witten, ``Quantum field theory and the jones polynomial,'' {\em
  Communications in Mathematical Physics}, vol.~121, no.~3, pp.~351--399, 1989.

\bibitem{leinaas1977theory}
J.~M. Leinaas and J.~Myrheim, ``On the theory of identical particles,'' {\em Il
  Nuovo Cimento B (1971-1996)}, vol.~37, no.~1, pp.~1--23, 1977.

\bibitem{Wilczek1982}
F.~Wilczek {\em Phys. Rev. Lett.}, vol.~49, pp.~957--959, 1982.

\bibitem{Frohlich1990}
J.~Fr{\"o}hlich and F.~Gabbiani {\em Reviews in Mathematical Physics}, vol.~2,
  no.~03, pp.~251--353, 1990.

\bibitem{Moore1991}
G.~Moore and N.~Read {\em Nuclear Physics B}, vol.~360, no.~2-3, pp.~362--396,
  1991.

\bibitem{wen1991non}
X.-G. Wen, ``Non-abelian statistics in the fractional quantum hall states,''
  {\em Physical review letters}, vol.~66, no.~6, p.~802, 1991.

\bibitem{Freedman2002}
M.~H. Freedman, M.~Larsen, and Z.~Wang {\em Communications in Mathematical
  Physics}, vol.~227, no.~3, pp.~605--622, 2002.

\bibitem{Freedman2003}
M.~Freedman, A.~Kitaev, M.~Larsen, and Z.~Wang {\em Bulletin of the American
  Mathematical Society}, vol.~40, no.~1, pp.~31--38, 2003.

\bibitem{Mong}
R.~S.~K. Mong, D.~J. Clarke, J.~Alicea, N.~H. Lindner, P.~Fendley, C.~Nayak,
  Y.~Oreg, A.~Stern, E.~Berg, K.~Shtengel, and M.~P.~A. Fisher {\em Phys. Rev.
  X}, vol.~4, p.~011036, 2014.

\bibitem{Hu2018}
Y.~Hu and C.~Kane {\em Physical review letters}, vol.~120, no.~6, p.~066801,
  2018.

\bibitem{Feiguin2007}
A.~Feiguin, S.~Trebst, A.~W. Ludwig, M.~Troyer, A.~Kitaev, Z.~Wang, and M.~H.
  Freedman {\em Physical review letters}, vol.~98, no.~16, p.~160409, 2007.

\bibitem{ludwig2011two}
A.~W. Ludwig, D.~Poilblanc, S.~Trebst, and M.~Troyer, ``Two-dimensional quantum
  liquids from interacting non-abelian anyons,'' {\em New Journal of Physics},
  vol.~13, no.~4, p.~045014, 2011.

\bibitem{affleck2017majorana}
I.~Affleck, A.~Rahmani, and D.~Pikulin, ``Majorana-hubbard model on the square
  lattice,'' {\em Physical Review B}, vol.~96, no.~12, p.~125121, 2017.

\bibitem{Kane2002}
C.~L. Kane, R.~Mukhopadhyay, and T.~C. Lubensky {\em Phys. Rev. Lett.},
  vol.~88, p.~036401, 2002.

\bibitem{Teo2014}
J.~C.~Y. Teo and C.~L. Kane {\em Phys. Rev. B}, vol.~89, p.~085101, 2014.

\bibitem{trebst2008short}
S.~Trebst, M.~Troyer, Z.~Wang, and A.~W. Ludwig, ``A short introduction to
  fibonacci anyon models,'' {\em Progress of Theoretical Physics Supplement},
  vol.~176, pp.~384--407, 2008.

\bibitem{buican2017anyonic}
M.~Buican and A.~Gromov, ``Anyonic chains, topological defects, and conformal
  field theory,'' {\em Communications in Mathematical Physics}, vol.~356,
  no.~3, pp.~1017--1056, 2017.

\bibitem{smirnov1991exact}
F.~Smirnov, ``Exact s-matrices for $\phi_{1, 2}$-perturbated minimal models of
  conformal field theory,'' {\em International Journal of Modern Physics A},
  vol.~6, no.~08, pp.~1407--1428, 1991.

\bibitem{mussardo1992off}
G.~Mussardo, ``Off-critical statistical models: factorized scattering theories
  and bootstrap program,'' {\em Physics Reports}, vol.~218, no.~5-6,
  pp.~215--379, 1992.

\bibitem{rowell2009classification}
E.~Rowell, R.~Stong, and Z.~Wang, ``On classification of modular tensor
  categories,'' {\em Communications in Mathematical Physics}, vol.~292, no.~2,
  pp.~343--389, 2009.

\bibitem{tsvelik2014integrable}
A.~Tsvelik {\em Physical review letters}, vol.~113, no.~6, p.~066401, 2014.

\bibitem{borcherding2018signatures}
D.~Borcherding and H.~Frahm, ``Signatures of non-abelian anyons in the
  thermodynamics of an interacting fermion model,'' {\em Journal of Physics A:
  Mathematical and Theoretical}, vol.~51, no.~19, p.~195001, 2018.

\bibitem{ji2019noninvertible}
W.~Ji and X.-G. Wen, ``Noninvertible anomalies and mapping-class-group
  transformation of anomalous partition functions,'' {\em Physical Review
  Research}, vol.~1, no.~3, p.~033054, 2019.

\bibitem{DiFrancesco1997}
P.~{Di Francesco}, P.~Mathieu, and D.~Senechal, {\em {Conformal field theory}}.
\newblock Springer, 1997.

\bibitem{majid2000foundations}
S.~Majid, {\em Foundations of quantum group theory}.
\newblock Cambridge university press, 2000.

\bibitem{frohlich2007duality}
J.~Fr{\"o}hlich, J.~Fuchs, I.~Runkel, and C.~Schweigert, ``Duality and defects
  in rational conformal field theory,'' {\em Nuclear Physics B}, vol.~763,
  no.~3, pp.~354--430, 2007.

\bibitem{davydov2011invertible}
A.~Davydov, L.~Kong, I.~Runkel, {\em et~al.}, ``Invertible defects and
  isomorphisms of rational cfts,'' {\em Advances in Theoretical and
  Mathematical Physics}, vol.~15, no.~1, pp.~43--69, 2011.

\bibitem{verlinde1988fusion}
E.~Verlinde, ``Fusion rules and modular transformations in 2d conformal field
  theory,'' {\em Nuclear Physics B}, vol.~300, pp.~360--376, 1988.

\bibitem{petkova2001generalised}
V.~Petkova and J.-B. Zuber, ``Generalised twisted partition functions,'' {\em
  Physics Letters B}, vol.~504, no.~1-2, pp.~157--164, 2001.

\end{thebibliography}

\end{document}